\documentclass[reprint, superscriptaddress, amsmath, amsfonts, aps, prb]{revtex4-2}
\usepackage{graphicx}
\usepackage{tabularx}

\begin{document}

\title{On Quantifying Large Lattice Relaxations in Photovoltaic Devices}

\author{Marco Nardone}
	\affiliation{Department of Physics and Astronomy, Bowling Green State University, Bowling Green, OH 43403, USA}
	\email{marcon@bgsu.edu}
\author{Yasas Patikirige}
	\affiliation{Department of Physics and Astronomy, Bowling Green State University, Bowling Green, OH 43403, USA}
\author{Kyoung E. Kweon}
	\affiliation{Lawrence Livermore National Laboratory, Materials Science Division, Livermore, California, 94550, USA}
\author{Curtis Walkons}
	\affiliation{Dept. of Mechanical Engineering, University of Nevada Las Vegas, Las Vegas, Nevada, 89154, USA}
\author{Theresa Magorian Friedlmeier}
	\affiliation{Zentrum für Sonnenenergie- und Wasserstoff-Forschung Baden-Württemberg, Stuttgart, Germany}	
\author{Joel B. Varley}
	\affiliation{Lawrence Livermore National Laboratory, Materials Science Division, Livermore, California, 94550, USA}
\author{Vincenzo Lordi}
	\affiliation{Lawrence Livermore National Laboratory, Materials Science Division, Livermore, California, 94550, USA}
\author{Shubhra Bansal}
	\affiliation{Dept. of Mechanical Engineering, University of Nevada Las Vegas, Las Vegas, Nevada, 89154, USA}	

\date{\today}

\begin{abstract}
Temporal variations of Cu(In,Ga)Se$_2$ photovoltaic device properties during light exposure at various temperatures and voltage biases for times up to 100 h were analyzed using the kinetic theory of large lattice relaxations. Open-circuit voltage and p-type doping increased with charge injection and decreased with temperature at low injection conditions. Lattice relaxation can account for both trends and activation energies extracted from the data were approximately 0.9 and 1.2 eV for devices with lower and higher sodium content, respectively. In these devices, increased sodium content resulted in higher initial p-type doping with greater stability. First principles calculations providing revised activation energies for the $(V_{Se}-V_{Cu})$ complex suggest that this defect does not account for the metastability observed here.

\end{abstract}
\maketitle


\section{Introduction}\label{sec:intro}

Large lattice relaxations (LLR) are changes in local lattice configurations around defects due to coupling between the electronic and atomic systems.  These defects are often metastable in that they can undergo thermally activated transitions driven by charge injection, photoexcitation and/or thermal energy and they can return to ground by thermal annealing \citep{redfield_photo-induced_2006}. Hence, the single defect level picture is insufficient and defect transformations can result in unique, time-dependent electronic properties.  Some examples include persistent photoconductivity (PPC) in III-V materials (such as, AlGaAs, caused by DX and EL2 center defects \citep{henry_nonradiative_1977, baraff_electronic_1992}), PPC in II-VI materials (such as, CdS \citep{wright_conductivity_1968, yin_large_2018} and CdTe \citep{lorenz_properties_1964}), photo-degradation of hydrogenated amorphous silicon (a-Si:H) caused by dangling bond defects \citep{staebler_reversible_1977}, charge-induced free carrier lifetime decay in crystalline silicon due to metastable boron-oxygen complexes \citep{bothe_fundamental_2005}, and impediments to doping of CdTe due to self-compensation by AX centers \citep{huang_nature_2015,nagaoka_high_2018}.  Fundamental understanding of materials with LLR defects can improve the performance and reliability of semiconductor devices.  In this work we present a general approach for quantifying activation energies associated with LLR using \emph{in situ} stress experiments, reaction kinetics analysis, and first principles calculations.  Copper indium gallium diselenide (CIGS) photovoltaic (PV) devices are considered as a specific application.

CIGS solar cells are commercially important with lab-scale power conversion efficiency surpassing 22\% \cite{green_solar_2018}.  PPC has been observed in polycrystalline CIGS films and attributed to LLR \citep{meyer_spectral_2002}. A related and common observation is the metastable drift of electronic properties over time when exposed to heat, light, and/or voltage bias \citep{ruberto_time-dependent_1987, igalson_metastable_1996, rau_persistent_1998, heise_light-induced_2017, nishinaga_effects_2017}. Typically, the net acceptor (hole) concentration, $N$, and open-circuit voltage, $V_{oc}$, increase with temperature under charge injection conditions, whereas dark annealing (without charge injection) tends to reduce $N$ and $V_{oc}$. Similar effects have been observed in single crystal CIGS PV devices \citep{nishinaga_single-crystal_2018}.

The microscopic nature of the metastable defect was originally thought to be a negative-$U$ center because it was found by deep level transient spectroscopy that two electrons were consumed (or holes formed) for every defect created \citep{igalson_metastable_1996}.  Later, \emph{ab-initio} calculations showed that the (V$_{Se}$-V$_{Cu}$) divacancy complex (with negative-$U$ properties) could exhibit LLR transitions between donor and acceptor configurations \citep{lany_light-and_2006}; support for this model has been widely cited \citep{igalson_understanding_2009, serhan_investigation_2011, burgelman_advanced_2013, obereigner_open-circuit_2015, maciaszek_quantitative_2018}. Experimental support \citep{urbaniak_creation_2009, okada_characterization_2011} for the calculated activation energies from Ref. \cite{lany_light-and_2006}  has been reported for time scales $< 1000$ s.  Our results over longer time scales of up to 100 hours indicate that a different defect species may be involved. As a first step, we provide revised first-principles energy calculations for the (V$_{Se}$-V$_{Cu}$) complex. Although the revised activation energies are mostly higher than the original ones, they are still rather low compared to those extracted from the data. Alternative defect species must be considered. 

An important property of high-efficiency CIGS devices is alkali content which is introduced either by sodium diffusion from the glass substrate or post-deposition treatment with NaF, KF, RbF, and/or CsF. Recently, it was shown that heat/light exposure causes greater acceptor formation in cells with KF \citep{nishinaga_effects_2017,khatri_impact_2018} and RbF \citep{ishizuka_effects_2018} than without alkali treatment. It appears that alkali impurities have an effect on LLR transition rates but further study is required. Herein we study CIGS cells with typical sodium content due to diffusion from the soda-lime glass substrate (Type 1) and reduced sodium due to a diffusion barrier (Type 2). 

\begin{figure}[htb]
\includegraphics[width=0.50\textwidth]{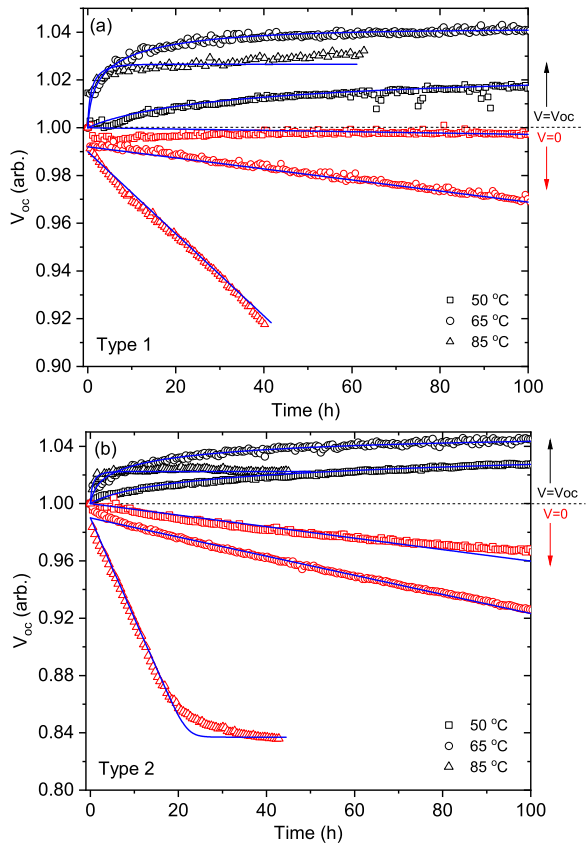}
\caption{Open circuit voltage ($V_{oc}$) normalized to the initial value as a function of time during stress at 0.1 W/cm$^2$ light intensity and the indicated voltage biases (0 or $V_{oc}$) and temperatures for (a) Type 1 and (b) Type 2 devices.  Points are data and lines are model fits from Eq. (\ref{eq:defects2}).\label{Fig:voc}}
\end{figure}

Our results for $V_{oc}$ as a function of time under light, voltage, and temperature stress are shown in Fig. \ref{Fig:voc}. A unique feature of this data is that performance metrics were measured \emph{in-situ} during stress at various temperatures and voltages, enabling a close examination of the time dependence. Most stress tests require intermittent removal of the device under study for measurements.  Metastability is evident in Fig. \ref{Fig:voc} with increasing $V_{oc}$ during high charge injection ($V=V_{oc}$ bias) and decreasing during low charge injection ($V=0$ bias); the rates of both processes increase with temperature. The solid lines are the LLR model fits, as described below.

\section{Reaction Kinetics}\label{sec:theory}

Variations in the quasi-Fermi levels (i.e. free carrier concentrations) due to external perturbations lead to changes in the occupancy (charge states) of defects.  Certain defects respond by a reorientation of the local the crystal lattice to lower the system energy. These structural relaxations are often thermally activated and can be described using the formalism of chemical reaction kinetics \cite{redfield_photo-induced_2006}. Defect reactions typically involve capture and/or emission of charge carriers; a single carrier process is a first-order reaction, simultaneous capture of two carriers is second order, and so on. The following subsections describe general first- and second-order reaction kinetics. In Sec. \ref{sec:results}, this approach is used to extract activation energies from the data.

\subsection{First-Order Kinetics}\label{sec:first}

If the LLR process is driven by electron capture and a first-order reaction is assumed, then the kinetics can be described by \cite{harju_electron-beam_2000},
\begin{equation}\label{eq:kinetics}
\frac{dN}{dt}=\alpha n - \beta N,
\end{equation}
where $N$ is the defect concentration (reaction product), $n$ is the \emph{excess} electron concentration, and $\alpha$ and $\beta$ are forward and backward reaction rates, respectively, with thermal activation energies $E_{\alpha}$ and $E_{\beta}$. $\alpha$ and $\beta$ are material properties that can also depend on local impurities. Note that electron capture can be interchanged with holes.

The excess electron concentration, $n$, depends on the defect density, $N$. If the defects are recombination centers, then the quasi-stationary approximation, $dn/dt=G-CNn\approx 0$ requires that $n=G/CN$, where $G$ is the generation rate and $C$ is the recombination coefficient. In the case of shallow acceptor defect formation, the mass action law requires that $n=n_i^2\gamma/N$, where $n_i$ is the intrinsic concentration and 
\begin{equation}\label{eq:gamma}
\gamma=\exp(eU/kT)-1
\end{equation}
defines electron injection through the quasi-Fermi level splitting $eU$.  In both cases, non-equilibrium charge injection invokes an effect to decrease the charge concentration toward equilibrium values, in accordance with Le Chatelier's principle.

Given an initial defect density of $N_0$ and the above expressions for $n(N)$, Eq. (\ref{eq:kinetics}) yields,
\begin{equation}\label{eq:defects}
N=N_{\infty}\sqrt{1-\left[1-\left(\frac{N_0}{N_{\infty}}\right)^2\right]\exp{\left(-2\beta t\right)}},
\end{equation}
with a saturation value of,
\begin{equation}\label{eq:sat1}
N_{\infty}=\sqrt{\alpha G/C\beta},
\end{equation}
for the case of recombination type defects, and
\begin{equation}\label{eq:sat2}
N_{\infty}=n_i\sqrt{\alpha\gamma/\beta},
\end{equation}
for shallow acceptor defects. The saturation level can be higher or lower than the initial concentration.  Under low injection conditions, $N_{\infty}\ll N_0$, and Eq. (\ref{eq:defects}) reduces to,
\begin{equation}\label{eq:anneal}
N=N_0\exp{\left(-t/\tau_a\right)}\quad\text{with}\quad\tau_a=1/\beta.
\end{equation}
Low injection can also be considered annealing conditions at elevated temperatures in the dark; hence the characteristic annealing time, $\tau_a$. For charge injection conditions but at relatively short times ($t\ll\tau_a$), before the onset of saturation, Eq. (\ref{eq:defects}) can be expressed as
\begin{equation}\label{eq:injection}
N=N_0\sqrt{1+t/\tau}\quad\text{with}\quad\tau=\frac{1}{2\beta}\left(\frac{N_0^2}{N_{\infty}^2-N_0^2}\right).
\end{equation}

\subsection{Second-Order Kinetics}\label{sec:second}

Reaction kinetics for the capture of two electrons can be described by,
\begin{equation}\label{eq:second}
\frac{dN}{dt}=\alpha n^2 - \beta N,
\end{equation}
with $\alpha$ having units of cm$^{-3}$ s$^{-1}$ in this case. Examples of double carrier processes include negative-$U$, $DX$, and $AX$ centers, which have been studied in several materials \cite{lany_intrinsic_2008, chadi_self-compensation_1989, wei_chemical_2002, yin_large_2018, henry_nonradiative_1977}.  From Eq. (\ref{eq:second}) and with the same $n(N)$ dependence described in Sec. \ref{sec:first}, defect evolution takes the form,
\begin{equation}\label{eq:defects2}
N=N_{\infty}\left\lbrace 1-\left[1-\left(\frac{N_0}{N_{\infty}}\right)^3\right]\exp{\left(-3\beta t\right)}\right\rbrace ^{1/3},
\end{equation}
with saturation limits for recombination-type defects and shallow acceptors as, respectively, 
\begin{equation}\label{eq:sat3}
N_{\infty}=(\alpha G^2/\beta C^2)^{1/3}
\end{equation}
and 
\begin{equation}\label{eq:sat4}
N_{\infty}=(\alpha n_i^4\gamma^2/\beta)^{1/3}.
\end{equation}

The annealing time, $\tau_a$, remains the same as in Eq. (\ref{eq:anneal}), but the charge injection case now has cube root time dependence,
\begin{equation}\label{eq:injection2}
N=N_0\left(1+t/\tau\right)^{1/3}\;\text{with}\;\tau=\frac{1}{3\beta}\left(\frac{N_0^3}{N_{\infty}^3-N_0^3}\right).
\end{equation}

Second order reactions can also occur when lattice relaxations are driven by the energy released during electron-hole pair recombination. In this case, the forward reaction is driven by the $np$ product \citep{stutzmann_light-induced_1985},
\begin{equation}\label{eq:recomb}
\frac{dN}{dt}=\alpha Anp - \beta N,
\end{equation}
where $p$ is the hole concentration and $Anp$ is the non-radiative recombination rate.  This model also produces $N\propto t^{1/3}$ behavior and was widely studied with respect to metastability in hydrogenated amorphous silicon \cite{staebler_reversible_1977}.

\section{Methods}\label{sec:exp}

\subsection{Device fabrication and characterization}\label{sec:fab}

Two sample types were evaluated: Type 1 (standard) and Type 2 (reduced sodium). The soda-lime glass substrate was cut, labeled, and cleaned prior to further processing. Type 2 samples had a sputtered AlO$_x$-AlN barrier layer (0.1 $\mu$m thick) to reduce the diffusion of sodium from the substrate.  The molybdenum back contact layer (0.05 $\mu$m thick) was sputtered in an inline process for all samples in the same run.  The subsequent CIGS layer (2 - 3 $\mu$m thick) was deposited in an inline coevaporation process and all samples were in the same CIGS run. The CdS buffer layer (0.05 $\mu$m thick) was deposited in batches by chemical bath deposition (CBD) according to standard procedures.  The buffer-window stack was CBD CdS followed by sputtered undoped ZnO and Al-doped ZnO layers (0.15 $\mu$m thick total). The cells were completed with NiAlNi grid contacts and the cell separation was by mechanical scribing. Each substrate contained 10 solar cell test structures. Average power conversion efficiencies were 16.2\% and 12.9\% for Type 1 and 2, respectively.  Details on device fabrication and performance are provided in Ref. \cite{nardone_baseline_2018}.

The Type 2 CIGS layer was grown under reduced sodium conditions, which influences the gallium and band gap gradients. The barrier applied in Type 2 cells was not as effective as had been demonstrated by previous tests. However, the expected trends of reduced efficiency and a flatter gallium gradient were apparent  \cite{nardone_baseline_2018}. According to secondary ion mass spectroscopy data (not shown), the approximate average sodium content was $10^{19}$ cm$^{-3}$ and $10^{18}$ cm$^{-3}$ for Type 1 and 2 devices, respectively.

\subsection{Characterization and Accelerated Stress Tests}\label{sec:stress}

$JV$ and $CV$ voltage sweeps were performed via a 4-probe setup.  $JV$ was measured with a Keithley 2400 source measure unit (SMU) in an ATLAS XXL+ chamber under dry conditions ($<15\%$ relative humidity), AM1.5G spectrum and 0.1 W/cm$^2$ illumination (Xe bulb source with spectral filters).  $CV$ was measured with a Solartron SI 1260 A, which is a combination frequency generator and impedance analyzer. The peak-to-peak AC voltage was 0.28 V and the frequency was 10 kHz to 1 MHz.  A resistance temperature detector (RTD) was placed on the device surface to monitor the temperature. Temperature control of cells at or near 25 $^\circ$C was provided by a Julabo recirculating water chiller/heater. 

Metastability was characterized by changes in $JV$ metrics determined by \emph{in-situ} current-voltage sweeps every 30 minutes under accelerated stress test (AST) conditions for a duration of 50-100 hours. The stress and measurement systems were integrated in the ATLAS XXL+ chamber.  Stress conditions were AM1.5G illumination of 0.1 W/cm$^2$ intensity, temperatures between 30 and 85 $^\circ$C, and voltage biases at open-circuit ($V=V_{oc}$) and short-circuit ($V=0$) conditions.  Pre- and post-stress \emph{ex-situ} $CV$ measurements were also performed immediately before, within a few hours, and within a few days after stress. All \emph{ex-situ} measurements were conducted at room temperature.

\subsection{Computational Details}\label{sec:comp}

Spin-polarized DFT calculations were performed with the Heyd-Scuseria-Ernzerhof (HSE06) screened hybrid functional \cite{heyd_hybrid_2003} and projector augmented wave (PAW) method \cite{blochl_projector_1994} as implemented in the VASP code \cite{kresse_efficiency_1996, kresse_efficient_1996}.  The plane-wave energy cutoff energy was set to 455 eV.  The fraction of Hartree-Fock exchange ($\alpha$) in the HSE functional was set to 30\%, which gives a bandgap of 1.13 eV for CuInSe$_2$, close to the experimental value \cite{moussa_band_2002}. The CuInSe$_2$ system was modeled using a large supercell containing 128 atoms. The standard supercell approach was employed to compute defect formation energies \cite{freysoldt_first-principles_2014}.  For charged defects, the electrostatic finite size error was corrected based on the correction scheme by Freysoldt, Neugebauer, and Van de Walle \cite{freysoldt_electrostatic_2011}. 

\section{Results}\label{sec:results}

\subsection{Quantifying Reaction Kinetics}\label{sec:quantify}

\begin{figure}[htb]
\includegraphics[width=0.45\textwidth]{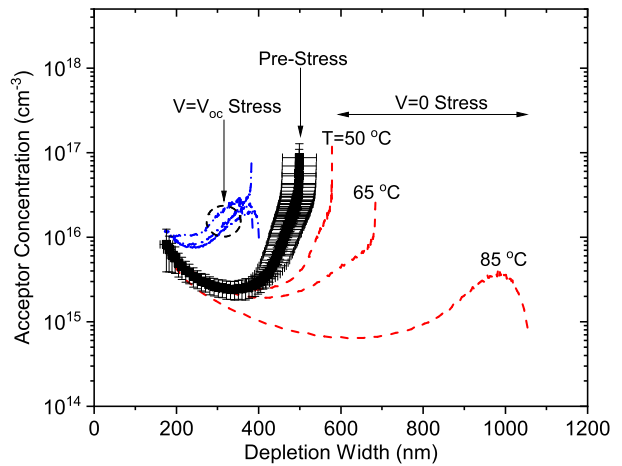}
\caption{Acceptor concentration versus depletion width for Type 1 devices: six before stress testing (black points plus/minus one standard deviation error bars), three stressed at $V=V_{oc}$ bias (dash-dot, blue lines), and three stressed at $V=0$ bias (dashed, red lines). Stress temperatures are shown for the $V=0$ bias stress. The average value of voltage bias for each cell is shown in Table \ref{tbl:stress}. \label{Fig:NvsW}}
\end{figure}

Light-, bias-, and heat-induced variations in doping are evident in Fig. \ref{Fig:NvsW} which shows the pre- and post-stress acceptor concentration ($N$) as a function of the p-n junction depletion width ($W$), (as typically derived from $CV$ data \cite{schroder_semiconductor_2006}) for six Type 1 devices (Type 2 results were similar but showed more drastic declines for the $V=0$ stress cases). Devices stressed at $V=0$ voltage bias (low charge injection) exhibited a trend of decreasing $N$ with temperature. Contrarily, an increase in $N$ was observed at $V=V_{oc}$ bias stress (high charge injection), but without apparent dependence on stress temperature.  In fact, $T=50 ^\circ$C stress led to a greater increase in $N$ than the higher temperatures. That may be indicative of the significant role of charge injection because although the cells were stressed at $V_{oc}$ bias, the magnitude of $V_{oc}$ (and therefore applied bias) decreased with temperature (by about 0.3$\%$/K for CIGS devices) and varied with time.  The time average value of $V_{oc}$ (bias stress, $\overline{V}$) at each temperature is shown in Table \ref{tbl:stress} along with the initial, $N_0$, and final, $N_f$, acceptor concentrations. Since $N$ varied with voltage, capacitance values from 0 to 0.5 V were used to determine the average $N_0$ and $N_f$, which provided a reasonable values of doping in the depletion region.

\begin{table}[htb]\centering
\caption{\label{tbl:stress}Stress temperature ($T$), average voltage ($\overline{V}$), and time ($t$) for Type 1 (typical sodium) and Type 2 (reduced sodium) devices.  Pre- ($N_0$) and post-stress ($N_f$) acceptor concentrations based on average of capacitance values from 0 to 0.5 V.}
	\begin{ruledtabular}
	\begin{tabular}{c c c c c c} 
		Type & $T$  & $\overline{V}$ & $t$  & $N_0$  & $N_f$  \\
		     &($^\circ$C) & ($mV$) & (h)& ($10^{15}$ cm$^{-3}$) & ($10^{15}$ cm$^{-3}$)\\
		\hline
		1  	 & 50  & 0 & 100 & 4.9 & 5.0  \\
		1  	 & 65  & 0 & 100 & 4.6 & 2.1  \\
		1  	 & 85  & 0 & 30 & 4.9 & 1.2  \\
		1  	 & 50  & 652 & 100 & 6.1 & 14.5  \\
		1  	 & 65  & 634 & 100 & 7.0 & 12.8  \\
		1  	 & 85  & 593 & 63 & 8.1 & 11.8  \\
		2  	 & 50  & 0 & 100 & 4.9 & 3.3  \\
		2  	 & 65  & 0 & 100 & 5.5 & 1.0  \\
		2  	 & 85  & 0 & 43 & 4.0 & 0.5  \\
		2  	 & 50  & 618 & 100 & 8.2 & 16.6  \\
		2  	 & 65  & 596 & 100 & 7.2 & 12.7  \\
		2  	 & 85  & 541 & 45 & 5.5 & 5.4  \\			
	\end{tabular}
	\end{ruledtabular}
\end{table}

Fig. \ref{Fig:NvsW} indicates clear variations in shallow acceptor concentration during light/bias/heat stress that correlates well with $V_{oc}$ drift.  Open circuit voltage as a function of time under stress up to 100 hours is shown in Figs. \ref{Fig:voc}(a) and (b) for cell Types 1 and 2, respectively.  Both cell types exhibited qualitatively similar behavior: increase in $V_{oc}$ when stressed at $V=V_{oc}$ bias (high charge injection) and decrease in $V_{oc}$ when stressed at short circuit conditions, $V=0$ conditions (low charge injection).  Similar to the acceptor density in Fig. \ref{Fig:NvsW}, the $V_{oc}$ reduction rate increased with temperature.  The $V_{oc}$ loss rate was clearly greater for Type 2 (reduced sodium) devices. Other metrics, including short-circuit current and fill factor, were relatively stable compared to $V_{oc}$.  Therefore, our analysis proceeds under the hypothesis that the LLR mechanism causes variations in shallow acceptors of density $N$ to an observable degree. Hence, the appropriate saturation limit is given by Eqs. (\ref{eq:sat2}) or (\ref{eq:sat4}).

The time dependence of $N$ was extracted from the $V_{oc}(t)$ data by noting that $V_{oc}\propto \ln{N}$. Fitting proceeded by first determining $\tau_a=1/\beta$ from the low injection $V=0$ stress cases by using Eq. (\ref{eq:anneal}) and the relation $V_{oc}(t)/V_{oc}(0)=\ln{N}/\ln{N_0}=1-\beta t/\ln{N_0}$ to establish a linear fit.  Initial values ($N_0$) were determined from the $CV$ data and are listed Table \ref{tbl:stress}.  From an Arrhenius plot of $\beta$, the activation energy, $E_\beta$, and pre-exponential, $\beta_0$ were determined.  Then, using the known value of $\beta$, the $V=V_{oc}$ early time stress data ($t\ll \tau_a$) were fit to determine $\tau$ (and therefore $\alpha$) using the logarithm of Eq. (\ref{eq:injection}) for first-order kinetics or Eq. (\ref{eq:injection2}) for second order kinetics.

\begin{figure}[htb]
\includegraphics[width=0.50\textwidth]{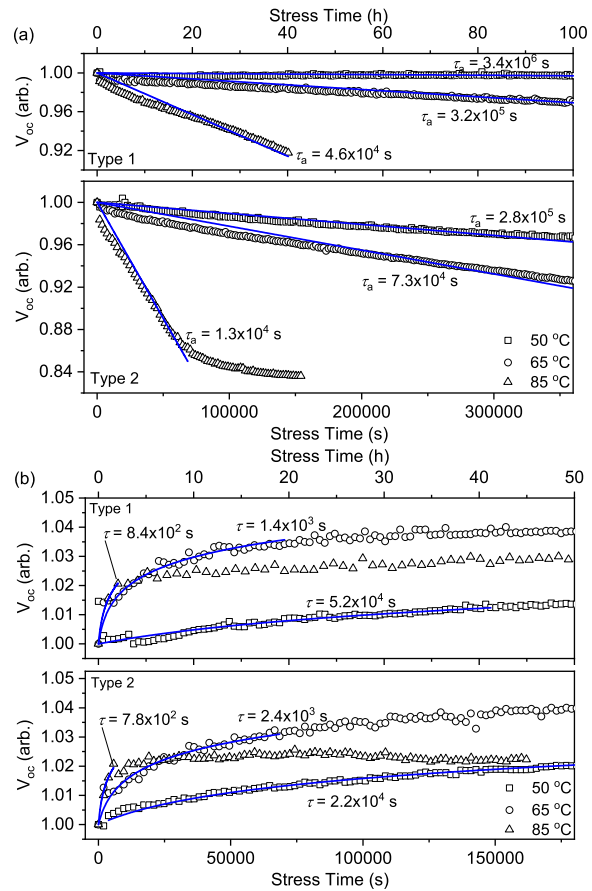}
\caption{Open circuit voltage with time under stress at 0.1 W/cm$^2$ light intensity and voltage of (a) $V=V_{oc}$ and (b) $V=0$ at the temperatures indicated.  The characteristic annealing time is shown for each curve fit (lines) using the logarithm of Eq. (\ref{eq:anneal}) for (a) and Eq. (\ref{eq:injection}) for (b). \label{Fig:fitting}}
\end{figure}

Fig. \ref{Fig:fitting}(a) shows that Type 2 devices tended to degrade more rapidly than Type 1 devices under $V=0$ light soak stress.  Fitting the data with Eq. (\ref{eq:anneal}) yielded characteristic annealing times on the order of 10$^4$ to 10$^6$ s for both types; Type 2 had shorter times.  Saturation was typically not observed in the $V=0$ bias cases, except for the Type 2 device at $T=85$ $^\circ$C, which saturated at about 15 hours. The high-injection, $V=V_{oc}$ stress cases shown in Fig. \ref{Fig:fitting}(b) indicate similar behavior and characteristic times for both device types.  Fitting with first- and second-order kinetic Eqs. (\ref{eq:injection}) and (\ref{eq:injection2}), respectively, provided similar results but second-order the $V=V_{oc}$ stress data slightly better.  Second-order fits are shown as the solid lines in Fig. \ref{Fig:fitting}(b).  Whether first- or second-order kinetics were prevalent cannot be inferred from this analysis; further discussion is provided in Sec. \ref{sec:disc}.

\begin{figure}[htb]
\includegraphics[width=0.45\textwidth]{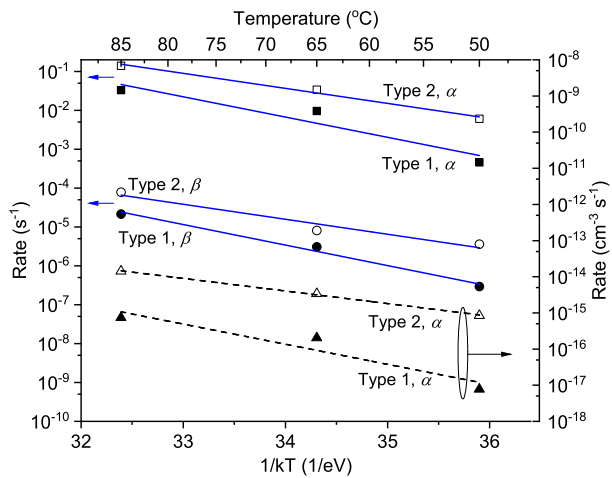}
\caption{Arrhenius plots of reaction rates $\alpha$ and $\beta$ for Type 1 and Type 2 devices.    Values for $\alpha$ were determined from first-order (fit with solid lines) and second order (fit with dashed lines) kinetics. Arrhenius fit values are provided in Table \ref{tbl:rates}. \label{Fig:arrhenius}}
\end{figure}

Arrhenius plots for the forward ($\alpha$) and backward ($\beta$) reaction rates are shown in Fig. \ref{Fig:arrhenius} and the extracted activation energies and pre-exponential values are listed in Table \ref{tbl:rates} for first- and second-order kinetics ($\beta$ is the same in both cases). Both forward and backward reaction rates are larger for Type 2 devices (reduced sodium). Although Fig. \ref{Fig:fitting}(b) suggests that Types 1 and 2 exhibited similar $V_{oc}$ increase, we note that Type 1 devices were subjected to higher voltage bias, as noted in Table \ref{tbl:stress}.  Figs. \ref{Fig:fitting} (a) and (b) indicate that sodium stabilizes doping in these CIGS devices.

\begin{table}[htb]\centering
\caption{\label{tbl:rates} Pre-exponential values and activation energies extracted from the Arrhenius plots for both device types.  (*) Units of $\alpha$ are (s$^{-1}$) for first-order and (cm$^{-3}$s$^{-1}$) for second-order kinetics. $\beta$ is order independent.} 	
	\begin{ruledtabular}	
	\begin{tabular}{c c c c c c}
	   Device & Kinetics  &  $\alpha_0$ & $E_{\alpha}$ & $\beta_0$ & $E_{\beta}$ \\
		Type &  Order     & (*)           & (eV)         &  (s$^{-1}$)  &  (eV)  \\
		\hline
		1 & 1 	 & $3.7\times 10^{15}$ & $1.20 \pm 0.36$  &  $3\times 10^{12}$ &   $1.22\pm 0.13$ \\
		2 & 1 	 & $5.0\times 10^{11}$ & $0.89 \pm 0.09$  &  $2\times 10^{8}$  &   $0.89\pm 0.19$  \\
		1 & 2 	 & $1.1\times 10^{3}$  & $1.28 \pm 0.39$  &  $3\times 10^{12}$ &   $1.22\pm 0.13$ \\
		2 & 2 	 & $2.8\times 10^{-3}$ & $0.80 \pm 0.04$  &  $2\times 10^{8}$  &   $0.89\pm 0.19$  \\
	\end{tabular}
	\end{ruledtabular}	
\end{table}

The above calculated reaction rates were also used with Eq. (\ref{eq:defects2}) to fit the entire range of $V_{oc}(t)$ data.  In Fig. \ref{Fig:voc}, the model curves are based on the pre-exponentials from Table \ref{tbl:rates} for second-order kinetics. The activation energies from fitting all of the curves with Eq. (\ref{eq:defects2}) had mean values of $E_{\alpha}=1.29$ eV and $E_{\beta}=1.21$ eV for Type 1 and $E_{\alpha}=0.81$ eV and $E_{\beta}=0.92$ eV for Type 2, with standard deviations of $<\pm 0.04$ eV. Although this was a qualitative (by eye) fitting exercise, it demonstrates that the entire range of data can be fit with reaction rates close to the values in Table \ref{tbl:rates}. Note that in Eq. (\ref{eq:gamma}) for $\gamma$, the Fermi level splitting is given by $eU\approx\overline{V}$ from Table \ref{tbl:stress} for the $V_{oc}$ bias cases. In the $V=0$ bias case, $\gamma$ is rather inconsequential for times prior to onset of saturation [cf. Eq. (\ref{eq:anneal})] and can be set to zero. However, for the Type 2 device stressed at $T=85$ $^\circ$C and $V=0$, saturation was observed and the model curve in Fig. \ref{Fig:voc}(b) was obtained with $eU=0.25$ eV, resulting in $N_{\infty}=2\times 10^{13}$ cm$^{-3}$. The value of $eU=0.25$ eV represents quasi-Fermi level splitting due to photo-excitation under short-circuit conditions.

\begin{figure}[htb]
\includegraphics[width=0.45\textwidth]{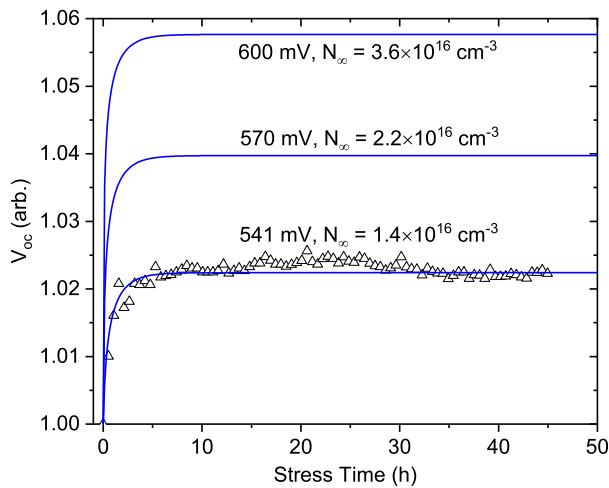}
\caption{$V_{oc}$ versus time for the Type 2 device stressed at 85 $^\circ$C and $V=V_{oc}$ (points). Eq. (\ref{eq:defects2}) and the relation $V_{oc}\propto \ln{N}$ at various applied bias are shown as curves along with the predicted saturation acceptor density, $N_{\infty}$. Input values are listed in Tables \ref{tbl:stress} and \ref{tbl:rates} (last row). \label{Fig:predict}}
\end{figure}

The apparent non-monotonic trend of saturation level with temperature in Fig. \ref{Fig:fitting}(b) is noteworthy.  Although the 85$^\circ$C, $V=V_{oc}$ stress case had the most rapid initial increase, it exhibited a lower saturation level than the 65$^\circ$C case.  That can be accounted for by the strong dependence of $N_{\infty}$ on Fermi level splitting, $eU\approx\overline{V}$, which was lowest for the 85 $^\circ$C case (see Table \ref{tbl:stress}). This point is illustrated in Fig. \ref{Fig:predict} where Eq. (\ref{eq:defects2}) was used to predict $V_{oc}$ evolution and saturation densities at 85 $^\circ$C for different voltage biases. The data for the Type 2 device is shown by the points and all of the model curves use the appropriate values from Tables \ref{tbl:stress} and \ref{tbl:rates} (last row of both tables).  A significant increase in $N_{\infty}$ (and $V_{oc}$) is predicted with voltage bias.

\subsection{First Principles Calculations}\label{sec:dft}

To understand the microscopic origin of the observed metastability, we revisited the ($V_{Se}-V_{Cu}$) divacancy in CuInSe$_2$.  As proposed earlier \cite{lany_light-and_2006}, the ($V_{Se}-V_{Cu}$) divacancy model in CIGS involves two predominant reactions due to electron ($e$) and hole ($h$) capture/emission \citep{lany_light-and_2006},
\begin{eqnarray}\label{eq:divacancy}
\left(V_{Se}-V_{Cu}\right)^+ + e\rightleftharpoons & \left(V_{Se}-V_{Cu}\right)^- + h\\
\label{eq:divacancy2}
\left(V_{Se}-V_{Cu}\right)^- + 2h\rightleftharpoons & \left(V_{Se}-V_{Cu}\right)^+.
\end{eqnarray}
The first reaction is electron capture ($ec$) in the forward direction and electron emission ($ee$) in reverse.  The second reaction is hole capture ($hc$) in the forward direction and hole emission ($he$) in reverse.  All four process are mediated by thermal activation energies

\begin{figure}[htb]
\includegraphics[width=0.47\textwidth]{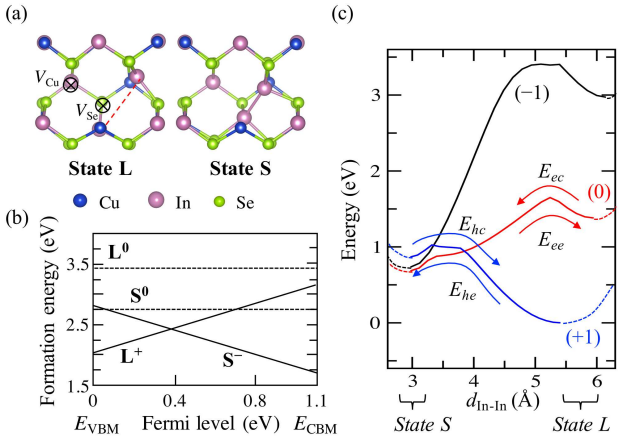}
\caption{(a) Optimized geometries of $V_{Se}-V_{Cu}$. (b) Defect formation energies of ($V_{Se}-V_{Cu}$) complex in CIS. (c) Configuration coordinate diagram of $(V_{Se}-V_{Cu})$ complex in its positive ($+1$), neutral ($0$), and negative ($-1$) charge-state. All energies are computed relative to the energy of the donor-type $(V_{Se}-V_{Cu})^+$ complex.  The respective relative energies correspond to $E_F$=$E_{VBM}$.  The parabolic curves were added (in dashed lines) beyond the equilibrium position for better understanding.  \label{Fig:DFT}}
\end{figure}

We computed the activation energies and the $(+/-)$ transition energy level of the ($V_{Se}-V_{Cu}$) complex for pure CuInSe (CIS) by using the HSE06 screened hybrid functional, which has been shown to yield a good description of the electronic and atomic structures and energetics in Cu(In,Ga)Se$_2$.  As shown in Fig. \ref{Fig:DFT}(a), there are two distinct local minimum configurations for the ($V_{Se}-V_{Cu}$) complex.  In the state L, two unsaturated In atoms are largely separated (denoted by the dashed line in Fig.\ref{Fig:DFT}(a)), while they form a dimer in the state S.  According to the formation energy calculations shown in Fig. \ref{Fig:DFT}(b),  the relative thermodynamic stabilities for these configurations vary depending on the charge state.  For the state L, the positive charge state (+) is always more favorable than the neutral state (0) and no $(+/0)$ charge transition level exists within the band gap. On the other hand, for the state S, the neutral state is the most stable when the Fermi level ($E_F$) is located very close to the valence band maximum (VBM), while the negative charge state $(-)$ becomes energetically more favorable than the neutral state when the Fermi level lies higher than 0.05 eV from the VBM.  Also, it should be noted that there is no additional charge transition level, such as $(-/2-)$, within the band gap.  This suggests that the state L and S serve as a shallow donor and acceptor, respectively. In Eqs. \ref{eq:divacancy} and \ref{eq:divacancy2}, the donor-type ($V_{Se}-V_{Cu}$)$^+$ and acceptor-type ($V_{Se}-V_{Cu}$)$^-$ correspond to the states L and S, respectively.  The donor to acceptor transition $(+/-)$ occurs at $E_F$=0.37 eV ($E_F$=0 corresponds to VBM), indicating that this transition would require the thermal activation energy associated with structural relaxation as well as the capture of charge carriers.  

\begin{table}[htb]
\caption{\label{tbl:Ea} Calculated energy barriers in eV from the configuration coordinate diagram in Fig. \ref{Fig:DFT}(c) and $(+1/-1)$ charge transition level $E_{tr}$. For reference, previously reported values \citep{lany_light-and_2006} are also listed.}
	\begin{ruledtabular}
	\begin{tabular}{c c c c c c}
		&   $E_{ec}$    &    $E_{ee}$   & $E_{hc}$ & $E_{he}$ & $E_{tr}$ \\
		\hline
		This work  &  0.27 	 & 0.97 & 0.16  &  1.03 &   0.37 \\
		Ref \cite{lany_light-and_2006}  &   $\sim$0.1 	 & 0.76 & 0.35  &  0.73  &   0.19  \\
	\end{tabular}
	\end{ruledtabular}
\end{table}

As shown in Fig. \ref{Fig:DFT}(c), we computed the configuration coordinate diagram of the $(V_{Se}-V_{Cu})$ complex in its positive ($+1$), neutral ($0$), and negative ($-1$) charge states to estimate the energy required for structural change between the donor- and acceptor-type configurations.  Starting from the lowest energy state, donor-type $(V_{Se}-V_{Cu})^+$, capturing one electron could convert it to the acceptor-type $(V_{Se}-V_{Cu})^0$ state.  The energy required for this process ($E_{ec}$) is calculated to be 0.27 eV as listed in Table \ref{tbl:Ea}.  As a shallow acceptor, the acceptor-type $(V_{Se}-V_{Cu})^0$ is likely ionized to $(V_{Se}-V_{Cu})^-$ by releasing one hole, which would increase the hole concentration as predicted in the reaction of Eq. (\ref{eq:divacancy}).  For the reverse reaction in Eq. (\ref{eq:divacancy2}), the acceptor-type $(V_{Se}-V_{Cu})^-$ can revert back to the donor-type $(V_{Se}-V_{Cu})^+$ by capturing simultaneously two holes and overcoming the energy barrier ($E_{hc}$) of  0.16 eV.  It should be mentioned that our computed energy barriers and $E_{tr}$ in Table \ref{tbl:Ea} are slightly different from those values reported earlier by Lany et al. \citep{lany_light-and_2006}.  The difference could be largely attributed to use of a different exchange-correlational functional that describes the electron-electron interaction in the system within the density functional theory calculations.  In \citep{lany_light-and_2006},  they employed a local density approximation (LDA) \citep{kohn_self-consistent_1965}, which gives inaccurate band gap of CIS and charge transition levels.  Although they applied necessary energy corrections to fix errors originating from the LDA, they could not fix the inherent limitation of overbinding \citep{van_de_walle_correcting_1999}.  Due to the overly favorable bonding interaction, the energy gained by forming the In$-$In dimer in the acceptor-type configuration is likely overestimated.  Indeed, in the positive charge state, the energy of the acceptor-type configuration is 0.8 eV less stable than that of the donor-type from our hybrid calculation. In comparison, the energy difference of 0.3 eV is much smaller with the LDA calculation.  Such a tendency would likely underestimate $E_{he}$, while overestimating $E_{hc}$, as demonstrated in Table \ref{tbl:Ea}. Based on the computed energy barriers in Table \ref{tbl:Ea}, it is expected that the donor to acceptor and the acceptor to donor conversions would be dominated by an electron capture process in Eq. \ref{eq:divacancy} and hole capture in Eq. \ref{eq:divacancy2}, respectively.

While we estimated accurate activation energies required for the transition between the donor-type and acceptor-type configurations by using hybrid DFT functionals, the values of $E_{ec}$ and $E_{hc}$ are still significantly smaller than what we obtained by fitting the experimental measurements (see Table \ref{tbl:rates} for $E_{\alpha}$ and $E_{\beta}$).  

\section{Discussion}\label{sec:disc}

The activation energy uncertainties in Table \ref{tbl:rates} are relatively large and could be improved by including additional measurements at each stress condition (work underway). We note that uncertainty is inherently large in this type of system.  For example, our results indicate that the activation energies are sensitive to sodium content, which is difficult to control during polycrystalline film deposition. Furthermore, LLR phenomena are often associated with dispersive reaction kinetics  due to distributions of activation energies \cite{shimakawa_persistent_1986, freitas_kinetics_2017}.

Although the kinetic models do not specify the microscopic nature of the defects, they provide important guidance for evaluation of candidate defect species by additional experiments and \emph{ab initio} calculations. Moving forward, the effects of alkali elements and oxygen on the $(V_{Se}-V_{Cu})$ divacancy and other defect complexes will be considered.

\section{Conclusions}\label{sec:conc}

The metastable behavior of CIGS PV devices was studied by means of \emph{in-situ} stress methods, reaction kinetics analysis, and first principles calculations.  Large lattice relaxations can account for both open-circuit voltage loss and gains depending on charge injection levels. Lattice relaxation activation energies extracted from the data were approximately 0.90 and 1.20 eV for devices with lower and higher sodium content, respectively. First principles calculations suggest that the $(V_{Se}-V_{Cu})$ complex is not responsible for the metastability observed here.

\section{Acknowledgements}

Thanks to Dr. Hajime Shibata and the Compound Semiconductor Thin Film team at AIST, Tsukuba, Japan for fruitful discussions.  This work was developed based upon funding from the Department of Energy Solar Energy Technology Office under award number DE-EE-0007750. Part of this work was performed under the auspices of the U.S. Department of Energy at Lawrence Livermore National Laboratory under Contract No. DE-AC52-07NA27344.


\bibliography{metastability_refs}

\end{document}